\documentclass[proceedings]{JHEP} 

\font\tenmsbm=msbm10 scaled 1200
\font\sevenmsbm=msbm9
\newfam\msbmfam
\textfont\msbmfam=\tenmsbm
\scriptfont\msbmfam=\sevenmsbm

\def\msbmn{\fam\msbmfam\sevenmsbm}
\def\beq{\begin{equation}}
\def\eeq{\end{equation}}
\def\bea{\begin{eqnarray}}
\def\eea{\end{eqnarray}}
\def\e{\epsilon}
\def\a{\alpha}
\def\da{\dot{\a}}
\newcommand{\dfrac}{\displaystyle \frac}

\conference{Quantum aspects of gauge theories, supersymmetry and unification.}

\title{Supergravity Predictions on Conformal Field Theories\thanks{Talk delivered by G. Dall'Agata.}}

\author{A.  Ceresole$^a$\thanks{Permanent Address: Dipartimento di Fisica, Politecnico di
 Torino, Corso Duca Degli Abruzzi 24, I-10129 Torino, Italy.},
G. Dall'Agata$^b$, R.  D'Auria$^{a\dag}$ and
 S. Ferrara$^a$
\\
$^a$  TH Division, CERN, 1211 Geneva 23, Switzerland. \\
$^b$ Dipartimento di Fisica Teorica, Universit\`a  di Torino,  \\
 Istituto Nazionale di Fisica Nucleare, Sezione di Torino,\\
via P. Giuria 1, I-10125 Torino.\\
E-mail: \email{dallagat@to.infn.it}}

\abstract{We give an update on recent results about the matching  between
CFT operators and KK states in the AdS/CFT correspondence, and add
some new comments on the realization of the baryonic symmetries from the
supergravity point of view.}


\begin{document}

\section{Introduction}

Over two years have passed since the proposal of Maldacena
\cite{M} of a correspondence between supergravity and string
theories on Anti de Sitter (AdS) spaces and conformal field
theories (CFT) on their boundary, in large $N$ limits. During this
period, this conjectured relation has been expressed more
precisely
 \cite{GKP,W}, it has been investigated under many
aspects, partially verified in various cases and also extended in
different directions \cite{MAGOO}.

One of the tests which has been carried out in great depth, giving
also some unexpected new results, is the matching of the spectra
of conformal operators on the CFT side with the Kaluza--Klein (KK)
excitations in the compactified supergravity. The AdS/CFT
correspondence indeed predicts a fixed relation between  scaling
dimensions and KK mass modes, which can be tested in many
examples. This matching was first proposed and used in \cite{W},
where it was called the  ``comparison to experiment" of the
AdS/CFT conjecture. In a first stage, it had been performed only
for the maximal supersymmetric cases (i.e. compactifications on
spheres) and for the lower supersymmetric models deriving from
orbifold compactifications \cite{MAGOO}.

Our group has extensively focused on the generalization of the
spectra matching test to lower supersymmetric models obtained by
supergravity compactifications on the product of AdS space with
various Einstein manifolds \cite{CDDF,BB,DF}. Due to the presence of
extra global symmetries inherited from the isometries of the
internal manifold, beside the $R$-symmetries, these models have a
far richer structure and thus yield much more probing proofs of
the AdS/CFT conjecture. In spite of the greater technical
complexity of lower (super)symmetric cases, we have chosen to
engage in their thorough study because we had at our disposal
quite powerful tools for supergravity analysis, such as harmonic
expansion on coset manifolds, that were developed in the old days
in the context of KK reduction of supergravity models
\cite{libro}.
 We
would like to collect here our main results and provide a brief
resum\'e of the lessons we have learned by exploring this subject.

\section{A test of the correspondence}

In the investigation of supergravity theories with lower
supersymmetry given by compactifications on coset manifolds, one
encounters  a very interesting and elaborate multiplet structure
which  makes possible some non--trivial checks in the
correspondence with the spectra of conformal operators of the
boundary field theory. In fact, differently from the spheres,
 where
all KK modes belong to short representations of supersymmetry and
thus have mass values that are protected against quantum
corrections, for less symmetric cosets one also  finds long and
semilong representations, that in principle do not have any
protection mechanism to prevent them from running with the
couplings. It is thus quite remarkable that one can nevertheless
establish a full map between each kind of KK multiplet and
appropriate families of conformal operators and their descendants.

We have essentially explored two directions: the correspondence
$AdS_5/CFT_4$ and $AdS_4/CFT_3$.

The $AdS_5/CFT_4$ case is more directly relevant from a physical
point of view, since it involves four dimensional gauge theories,
but also the $AdS_4/CFT_3$ one offers some intriguing challenges
which could give us more insight in the formulation of the
conjecture.

For spontaneous  compactifications of type IIB supergravity on a
five dimensional coset manifold,  there is only one space
preserving some supersymmetry\cite{Rom}: $$T^{11} \equiv
\dfrac{SU(2) \times SU(2)}{U(1)}$$ where the $U(1)$ factor is
embedded diagonally in the two $SU(2)$. We have determined the
full KK spectrum on this manifold \cite{CDD} (extending some
previous partial results \cite{G,JR}) and then tested  the AdS/CFT
map \cite{CDDF}, by matching it against the spectrum of primary
conformal operators of the dual CFT constructed in \cite{KW}. In
this example we have not only shown that the duality works, but we
have also  some new hints on the CFT behaviour.

The extension of such study to $M$--theory compactifications on
seven--manifolds is much more complicated. It is indeed known that
three dimensional CFT's are difficult to analyze because they
emerge in non--perturbative limits of conventional gauge field
theories. Moreover, if for type IIB on $T^{11}$ we had a well
defined CFT to be used towards the comparison, for the $M$--theory
compactifications  a well established CFT was not available. Thus
we have  {\sl used} the correspondence, at first  to guess these
CFT's, and then to verify by matching the spectra whether they
were well--defined.

$M$--theory allows a variety  of supersymmetric compactifications
down to four dimensions. The ${\cal N} = 2$ examples  can be
divided  into two categories \cite{D}: toric ones $$M^{111}
\equiv \dfrac{SU(3) \times SU(2) \times U(1)}{SU(2) \times U(1)
\times U(1)} \quad \hbox{and }$$ $$Q^{111} \equiv \dfrac{SU(2)
\times SU(2) \times SU(2)}{U(1) \times U(1)}$$ and non--toric
ones $$V_{(5,2)} \equiv \frac{SO(5)}{SO(3)}.$$ While for the
first, toric geometry helps in the definition of the CFT
\cite{BB}, in the non--toric case one has to deal with even harder
difficulties \cite{CDDF2}.

\medskip

Summing up, in this analysis we have met three main features which
are worth describing in some detail: i) the agreement between the
CFT expectations and the supergravity results, ii) the existence
of  long multiplets with rational energy quantum numbers predicted
by supergravity, and iii) the identification of the baryonic
symmetries as those deriving from the well known
 presence of Betti multiplets \cite{Betti} in
the compactified supergravity.

\section{Matching the spectra}

The $AdS/CFT$ correspondence can be used in two ways:  either to
control the validity of the CFT by predicting properties of the
supergravity, such as the mass spectrum, or to obtain  information
from tree level calculations in supergravity on the strong
coupling CFT behaviour.

Not only the fixed relation required by the $AdS/CFT$ map  between
the anomalous dimension of the various boundary conformal fields
and the masses of the bulk  KK modes holds for lower supersymmetry
as for the highest symmetric cases \cite{CDDF,BB,CDDF2}, but there
exists a full correspondence between {\it all} the KK modes and
the conformal operators of preserved scaling dimension.

In order to give a taste of how this works, we turn to the
simplest non--trivial example, that is type IIB compactification
on $AdS_5 \times T^{11}$.

The dual four--dimensional $CFT$ was given in \cite{KW} as an
${\cal N}=1$ Yang--Mills theory with a flavor symmetry
$G=SU(2)\times SU(2)$. It should describe the physics of a large
number ($N$) of $D3$--branes placed at the singular point of the
cone over the $T^{11}$ manifold in the decoupling limit.

The ``singleton'' degrees of freedom of the CFT, called  $A$
and $B$, are each a doublet of the $G$ factor groups
and have a conformal anomalous dimension $\Delta_{A,B}=3/4$.
The gauge group ${\cal G}$ is $SU(N)\times SU(N)$ and the $A$ and $B$
chiral multiplets transform in the $(N,\overline N)$ and
$(\overline N,N)$ of ${\cal G}$ respectively.
The gauge potentials lie in the adjoint of one of the two $SU(N)$
groups, and their field--strength in superfield notation is given by $W_\alpha$.
They are singlet of the matter groups, with $R$--symmetry charge $r=1$
and $\Delta = 3/2$.

There is also a superpotential given by  \cite{KW} \beq {\cal W} =
\lambda \e^{ij} \e^{kl} \,  Tr(A_i B_k A_j B_l),\eeq which has
$\Delta = 3$, $r = 2$. It plays an important role in the
discussion in that it determines to some extent both the chiral
spectrum and the marginal deformations of the SCFT.

Knowing the fundamental degrees of freedom of the conformal field
theory, one could try to write the conformal operators by simply
combining the above fields into all possible products while
respecting the symmetries of the theory.

The first operators one can build in this way are the chiral
operators \beq \label{TRAB} Tr(AB)^k \eeq which are those with the
lowest possible dimension for a given $R$--charge (they have
indeed $\Delta = \frac{3}{2} r = \frac{3}{2}k$). We notice that in
the (\ref{TRAB}) operators we can freely permute all the $A$'s and
$B$'s by using the equations for a critical point of the
superpotential
\beq B_1\ A_k\ B_2=B_2\ A_k\ B_1\ , A_1\ B_l\
A_2=A_2\ B_l\ A_1\ .
\eeq

Next to these, one could also have an operator given by
$Tr[W_\alpha (AB)^k]$ or $Tr[W^2 (AB)^k]$, and so on. But which
are the operators with protected dimension? This is a crucial
question, since only the protected operators find a matching state
among the KK fields, while those that suffer from quantum
corrections are to be found within the full string theory.

It is a well--established result that operators with protected conformal
dimension correspond to the short representations of the supergroup
which they belong to.

For $T^{11}$, this supergroup is $SU(2,2|1)$ , while for the
previously mentioned $M$--theory cases it is  $OSp(4|2)$. More
generally, for $N$ supersymmetries the four dimensional case
involves $SU(2,2|{\cal N})$ whose shortening conditions in terms
of superfields have been explained in \cite{FZS}, while the
generic three--dimensional case involves $OSp(4|{\cal N})$ whose
shortenings have been recently discussed in \cite{FERR}.

In the $T^{11}$ example ($\a$, $\da$ are spinor indices. $x$,$\theta$
and $\bar{\theta}$ are the bosonic and fermionic coordinates)
we have only three types of such operators,
namely the {\it chiral}
\beq
\label{chiral}
\bar D^{\dot\alpha}S_{\alpha_1...\alpha_{2J}}=0,
\eeq
{\it conserved}
\bea
\label{cons}
 \bar D^{\dot\alpha_1}J_{\alpha_1...\alpha_{2J_1},
\dot\alpha_1...\dot\alpha_{2J_2}}&=& 0 \\ \hbox{and }
D^{\alpha_1}J_{\alpha_1...\alpha_{2J_1},\dot\alpha_1...\dot\alpha_{2J_2}}&=&0
\eea and {\it semi--conserved} superfields \bea \label{op} \bar
D^{\dot\alpha_1} L_{\alpha_1...\alpha_{2J_1},
\dot\alpha_1...\dot\alpha_{2J_2}} (x,\theta,\bar\theta)&=&0, \\
\nonumber (\bar D^2L_{\alpha_1...\alpha_{2J_1}}=0\, \hbox{for}\,
J_2=0). \eea These differential constraints imply that these
fields satisfy certain specific restrictions on their quantum
numbers. As a consequence,  their anomalous dimension is fixed in
terms of their spin and $R$--symmetry charge. These constraints
are respectively: \beq r = \dfrac{2}{3} \Delta, \eeq for chiral
ones, \beq r = \dfrac{2}{3} (\Delta-2-2J_2) \eeq for semiconserved
ones and \beq
\begin{array}{rcl}
r &=& \dfrac{2}{3} (J_1 - J_2),\\
\Delta &=& 2 + J_1 + J_2,
\end{array}
\eeq
for conserved ones.

It is easy to relate operators of different type by superfield
multiplication. The product of a chiral $(J_1,0)$ and an anti--chiral $(0,J_2)$
primary gives a generic superfield with
$(J_1,J_2)$, $\Delta = \Delta^c +
\Delta^a$ and $r = \frac{2}{3}(\Delta^c - \Delta^a)$.
By multiplying a {\it conserved current} superfield $J_{\a_1 \ldots
\a_{2J_1}, \da_1 \ldots \da_{2J_2}}$ by a chiral scalar
superfield one gets a semi--conserved superfield with $\Delta =
\Delta^{c} + 2 + J_1 + J_2$  and $r = \frac{2}{3}(\Delta-2-2J_2)$.

These are the basic rules to construct operators with protected
dimensions beside the chiral ones, and they also apply in superconformal
field theories of lower or higher dimensions.

\medskip

Since the anomalous dimensions of the protected operator is fixed
in terms of
their spin and $R$--symmetry, it must be given by a rational number. This
condition severely
restricts the search for the corresponding
supergravity states, as it  imposes strong constraints on
the allowed masses and matter group quantum numbers.

We find in our analysis that the requirement for
the anomalous dimensions to be rational implies that one must look for
dual  states also having  rational masses .

The virtue of KK harmonic analysis on a coset space hinges on the
possibility of reducing the computation of the mass eigenvalues of
the various kinetic differential operators to a completely algebraic
problem, while it allows to eliminate completely any
explicit dependence on the coordinates of the internal manifold.
Harmonics are uniquely identified by $G$ quantum numbers, and they are acted
upon by derivatives that are reduced to algebraic operators.
Such elegant technique can be quite cumbersome for complicated cosets
\cite{BB,CDDF2}, but it is quite straightforward for the simple
$T^{11}$ manifold, where it leads beyond the computation of
the scalar laplacian eigenvalues \cite{G}, or of specific sectors of
the mass spectrum \cite{JR}.

By diagonalizing different operators for fields of various spin,
we have found that all the masses have a fixed dependence on the
scalar laplacean eigenvalue \beq H_0(j,l,r)=6[j(j+1)+l(l+1)-1/8
r^2] \label{accazero} \eeq where $(j,l,r)$ refer to the
$SU(2)\times SU(2)$ and to the $R$--symmetry quantum numbers.

This gives us a new element in the analysis as we will soon see, since
besides the $SU(2,2|1)$ quantum numbers, we have also to match those of the
matter group.

The full analysis \cite{CDD} reveals that the supergravity theory has
one long graviton multiplet with conformal dimensions
\beq
\Delta = 1+\sqrt{H_0(j,l,r)+4},
\eeq
 four long gravitino multiplets with
\beq
\begin{array}{rcl}
\Delta &=& -1/2 + \sqrt{H_0(j,l,r\pm 1) + 4}, \\
\Delta &=& 5/2 + \sqrt{H_0(j,l,r\pm 1) + 4},
\end{array}
\eeq
and four long vector multiplets, with
\bea
\Delta = -2 + \sqrt{H_0(j,l,r)+4}, &&\nonumber\\
\Delta = 4 + \sqrt{H_0(j,l,r)+4}, &&\\
 \Delta = 1 + \sqrt{H_0(j,l,r\pm 2)+4}. && \nonumber
\eea
Beside these long ones, there are the shortened supermultiplets.

The above formulae clearly show that the
conformal dimensions become rational when the square roots
assume rational values
\beq
H_0+4 \in \hbox{\msbmn Q}^2.
\eeq

This equation  is found to admit some special solutions for
\bea
&j=l=|r/2|,&\\
&j=l-1=|r/2| \; {\rm or} \; l=j-1=|r/2| .&
\eea
Given these strong constraints on the possible
$SU(2,2|1)$ quantum numbers as well as on the $SU(2) \times SU(2)$
ones, it becomes an easy task to build the appropriate conformal operators
satisfying such constraints and find the relevant bulk
supermultiplets.

While referring to \cite{CDDF} for all details,
we list some interesting examples of conformal operators.

The chiral operators of the conformal field theory are given by
\bea
\label{prim}
S^k &=& Tr (AB)^k \\
\label{sec}
\Phi^k &=& Tr \left[ W^2 (AB)^k \right]\\
\label{ter}
T^k &=& Tr \left[ W_\alpha (AB)^k \right]
\eea
and are shown to correspond to hyper--multiplets containing massive recursions
of the dilaton or the internal metric (\ref{prim} and \ref{sec})
or to tensor multiplets (\ref{ter}).

Even more interesting are the towers of operators associated to the
semi--conserved currents.
Some of them are
\bea
{J}_{\a\da}^k &=& Tr(W_\alpha e^V \bar{W}_{\dot{\alpha}}e^{-V}(AB)^k), \\
{J}^k &=& Tr(Ae^V \bar{A} e^{-V} (AB)^k),
\eea
which lead  to short multiplets whose highest state is
a spin 2 and spin 1 field respectively, with masses given by
\bea
M_{J_{\alpha\dot{\alpha}}^k} = \sqrt{\frac{3}{2} k \left( \frac{3}{2} k +
4\right)},
\\
\displaystyle
\hbox{ and } \quad  M_{J^k} = \sqrt{\frac{3}{2} k \left( \frac{3}{2} k +
2\right)}.
\eea
These bulk states correspond to massive recursion of the graviton
and of the gauge bosons of the matter groups.

It has been explained that under certain conditions the
semi--conserved superfields can become conserved, and this is indeed the
case.
If we set $k=0$ we retrieve the conserved currents related
to the stress--energy tensor and the matter isometries .
In fact $M_{J_{\alpha\dot{\alpha}}^0} = M_{J^0} = 0$ are the
massless graviton and gauge bosons of the supergravity
theory.

The above analysis can be carried out for  $M$--theory compactifications,
where again a full
correspondence can be established for the {\it short} operators on the
CFT  and the {\it short} multiplets of the supergravity theory.
We must say however that, while in the $T^{11}$ case the superpotential
gives us a rule to discard all the sets of operators which
are not related to any KK state, for the $M$--theory KK spectra to
agree with the CFT operators one has to uncover some unknown quantum
mechanism \cite{BB} or the existence of some highly non trivial
superpotential \cite{CDDF2} that would eliminate the mismatching states.

Up to now
we have  checked the AdS/CFT correspondence as far as what the conformal
field theory imposes on the bulk states, but what can we learn {\sl
on the CFT}
from the analysis of the supergravity states?

\section{Supergravity predictions}

There are essentially two aspects of the supergravity theory which
can give us new insight on the dual CFT. The first is the presence
of long multiplets that nevertheless have  rational scaling
dimensions, which could provide us with new non--renormalization
theorems (at least in the large $N$, $g_s N$ limit). The other is
is the existence of the so--called Betti multiplets, which give
rise to additional symmetries in the boundary theory.

Let us now turn to the first aspect.

We have shown that the conformal operators with protected
dimension are given by chiral ones or by their products with the conserved
currents.
The surprising output of the supergravity analysis
is that there exist some conformal operators  that in spite of not being
protected by supersymmetry, still have rational conformal dimension.

Confining ourselves to the $T^{11}$ case, if we take the chiral
operator $$Tr( W^2 (AB)^k),$$ we can make it non--chiral by simply
inserting into the trace an antichiral combination of the gauge
field--strength $$Tr( W^2 e^V \bar{W}^2 e^{-V} (AB)^k).$$ This
operator then corresponds to a long multiplet in the bulk theory
and one should expect its scaling dimension to be generically
renormalized to an irrational number. If we search for the
corresponding vector multiplet in the supergravity theory, we see
that its anomalous dimension is instead rational and {\it matches
exactly the naive sum of the dimensions of the operators inside
the trace}. We find this  to be the case for all the lowest
non--chiral operators of general towers with irrational scaling
dimension. For instance, the  towers of operators \bea Tr\left[
W_\alpha (A e^V \bar A e^{-V})^n (AB)^k\right] \\ Tr\left[ e^V
\bar W_{\dot{\alpha}} e^{-V} (A e^V \bar A e^{-V})^n (AB)^k\right]
\eea have an irrational value of $\Delta$ for generic $n$, but
when $n = 1$ we have found that they do have rational anomalous
dimension $\Delta = 5/2 + 3/2 k$. When $n=0$ we retrieve the
chiral, or semi--conserved operators with protected $\Delta$. This
is a highly non--trivial prediction of the correspondence on the
CFT which comes only from the computation of the spectrum on the
KK side and we hope it could receive in the future an explanation
from the CFT point of view.

\medskip

If we restrict our attention to the protected operators, we could
say that the above peculiar feature arises also in the
$AdS_4/CFT_3$ case. However, we have a true one--to--one map and
full agreement on the two sides only for a specific
seven--dimensional compactification, that is the Stiefel manifold
$SO(5)/SO(3)$ \cite{CDDF2} (see the summary table therein).The
latter seems to be rather different from the other ${\cal N}=2$
compactifications  of \cite{BB}. Indeed, although  the spectra
look very similar, it seems that in the  examples dealt with in
\cite{BB}, for some of the supergravity states it is  not easy to
identify the related CFT operator.

\section{Betti multiplets}

The second AdS prediction on the CFT is the
existence of baryon symmetries.

As pointed out by Witten \cite{WittBaryon}, the existence of such
baryon symmetries is related to non--trivial Betti numbers of the
internal manifold.
Moreover, from the supergravity point of view, the non trivial value
of such numbers implies the appearance of extra massless multiplets, the
Betti multiplets \cite{CDD}.
It is then quite natural to propose a relation between the existence of
Betti multiplets and of baryon symmetries.

Let's see how this works.

The non--trivial $b_2$ and $b_3$ numbers of the $T^{11}$ manifold
imply the existence of closed non--exact 2--form $Y_{ab}$ and 3-form $Y_{abc}$.
These forms must be singlets under the full isometry group, and thus
they signal the presence of new additional massless states in the
theory than those connected to the $SU(2) \times SU(2) \times
U_R(1)$ isometry.

From the KK expansion of the complex rank 2 $A_{MN}$ and real rank
4 $A_{MNPQ}$ tensors of type IIB supergravity we learn that we
should find  in the spectrum a massless vector (from $A_{\mu
abc}$), a massless tensor (from $A_{\mu\nu ab}$) and two massless
scalars (from the complex $A_{ab}$).
This implies the existence of the so called Betti vector, tensor
and hyper--multiplets, the last two being a peculiar feature of
the $AdS_5$ compactification \cite{CDD}.
The additional massless vector can be seen to be the massless
gauge boson of an additional $U_B(1)$ symmetry of the theory.

From the boundary point of view we need now to find an
operator counterpart for such vector multiplet and look for an
interpretation of the additional symmetry.
The task of finding the conformal operator is very easy, once we
take into account that it must be a singlet of the full isometry
group and must have $\Delta = 3$.
The only operator we can write is \cite{CDDF,KW2}
\beq
\begin{array}{l}
{\cal U}=Tr\left(A e^V \bar A e^{-V}\right) - Tr\left(B e^V \bar
B e^{-V}\right) \\
(D^2 {\cal U}=\bar D^2 {\cal U}=0),
\end{array}
\eeq
which represents the conserved current of a baryon symmetry of
the boundary theory under which the $A$ and $B$ field transform
with opposite phase.
We have shown that the occurrence of such Betti multiplets is
indeed due to the existence of non--trivial two and three--cycles
on the $T^{11}$ manifold.
This implies that, from the stringy point of view, we can wrap the
$D3$--branes of type IIB superstring theory around such 3--cycles
and the wrapping number coincides with  the baryon number of
the low--energy CFT \cite{KW2}.

We would like to point out  that this feature of some manifolds
can be used to check the right dimension of the singleton fields as
done in \cite{BB}.
One can indeed compute the conformal dimension of the CFT operator
coupling to the baryon field obtained by a $Dp$--brane wrapping a
non--trivial $p$--cycle and match it with its mass, which should be
proportional to the volume of the same  cycle.

\section{A puzzle}

An interesting case where the baryonic symmetry does not appear to
be  simply related to the Betti multiplets is that of type IIA
compactification on $AdS_{4}\times \hbox{\msbmn P}^{3}$. This
gives a supergravity theory which should be dual to an ${\cal N}=
6$ CFT in three dimensions. It has been conjectured that the
supergravity spectrum should be the same for $M$--theory on
$AdS_{4}\times S^{7}/{\hbox{\msbmn Z}}_k$ (for $k> 3$) and for the
Hopf reduction of $AdS_{4}\times S^{7}$ on $AdS_{4}\times
\hbox{\msbmn P}^{3}$ \cite{P3}. It can indeed be shown
\cite{CDDF3} that the surviving states of the $M$--theory
expansion on $AdS_{4}\times S^{7}/{\hbox{\msbmn Z}}_k$ are the
same as those of the ${\cal N}= 8$ theory which are neutral under
the $U(1)$ along which we Hopf reduce $S^{7}$ to $\hbox{\msbmn
P}^{3}$. These are exactly the same as those appearing in the
harmonic expansion of type IIA theory on $AdS_{4}\times
\hbox{\msbmn P}^{3}$.

From these facts, one should deduce that the massless vector of
the additional $U(1)$ baryon symmetry is simply the KK vector
deriving from the reduction of the eleven dimensional metric on
the ten dimensional space $AdS_4 \times  \hbox{\msbmn P}^{3}$. But
here comes the puzzle.

Type IIA theory has a three--form $C$ which should give rise to Betti
vector multiplets when expanded on the internal manifold $\hbox{\msbmn P}^{3}$.
The complex projective space $\hbox{\msbmn P}^{3}$ has indeed a
non--trivial Betti two--form: the complex structure $J_{ab}$.
This implies that the expansion of
$C_{\mu ab}(x,y)$ in terms of the
harmonics of the internal manifold contains a
vector $c_{\mu}^{0}$ coupled to this form:
\beq
\label{sviluppo}
C_{\mu ab}(x,y) = \sum_{I} c_{\mu}^{I}(x) Y_{ab}^{I}(y) +
c_\mu^{0}(x) J_{ab}.
\eeq

This again could be interpreted as the massless vector of the baryon
symmetry, but we know we have only one such vector.

The solution lies in the fact that this  $c_{\mu}^{0}$ is
non--physical.
It is actually a pure gauge mode as we will shortly see.

Usually, type IIA supergravity is described by a one--form $A$, a
two--form $B$ a three--form $C$ and a dilaton $\Phi$ with
field--strengths:
\bea
F &=& dA,\\
H &=& dB, \\
G &=& dC + A \, dB.
\eea

If we define \beq C^\prime \equiv C + AB, \eeq then $dC^\prime =
dC + A\ dB- dA\ B$ and the four--form definition becomes \beq G =
dC^\prime + FB. \eeq At this point $G$ is trivially invariant
under \beq \delta C^\prime = d K, \hbox{ and } \left\{
\begin{array}{rcl} \delta A &=& d\Lambda \\ \delta C^\prime &=& 0
\end{array} \right., \eeq while $\delta B = d \Sigma$ requires
$\delta C = F \Sigma$. This implies that the physical invariance
of $B_{\mu\nu}(x)$, $\delta B_{\mu\nu}(x) = 2\partial_{[\mu}
\Sigma_{\nu]}(x)$ requires $C^\prime_{\mu ab}$ to transform
according to
\beq
\label{cpri}
\delta C^{\prime}_{\mu ab}(x,y) = F_{ab}
\Sigma_\mu.
\eeq
Keeping only linear terms in (\ref{cpri}), we get \beq \delta
C^\prime_{\mu ab} (x)  = J_{ab} \Sigma_\mu (x), \eeq which tells
us, by comparison with (\ref{sviluppo}) now applied to $C^\prime$,
that the generic mode $c_\mu^I$ is invariant $\delta_{\Sigma}
c_\mu^I = 0$, while $\delta_{\Sigma} c_\mu^0 = \Sigma_\mu(x)$ is a
pure gauge field.

\paragraph{Acknowledgements.}
We are glad to thank the local organizers for a stimulating conference and
their very warm hospitality.
S.F. is supported by the DOE under grant DE-FG03-91ER40662, Task C.
This work is also supported by the European Commission TMR programme
ERBFMRX-CT96-0045 (University and Politecnico of Torino and Frascati
nodes).

\end{document}